\documentclass{aa}
\usepackage{graphics,epsfig}

\newcommand{\kms}{km~s$^{-1}$}

\begin{document}

\title{Kinematics of the faintest  gas rich galaxy in the  Local Group: DDO210}
\titlerunning{Kinematics of DDO210}
\author{Ayesha Begum\inst{1}\thanks{ayesha@ncra.tifr.res.in} and 
        Jayaram N. Chengalur\inst{1}\thanks{chengalur@ncra.tifr.res.in}
}
\authorrunning{Begum \& Chengalur}
\institute{National Centre for Radio Astrophysics,
	Post Bag 3, Ganeshkhind, Pune 411 007}
\date{Received mmddyy/ accepted mmddyy}
\offprints{Ayesha Begum}
\abstract{ We present deep, high velocity resolution ($\sim 1.6$ km sec$^{-1}$)
Giant Meterwave Radio Telescope (GMRT) HI 21cm synthesis images of the 
faint ($M_B \sim -10.6$) local group dwarf galaxy DDO210.
We find that  the HI distribution in the galaxy is not axi-symmetric,
but shows density enhancements in the  eastern and southern halves of 
the galaxy. The optical emission is lopsided with respect to the HI,
most of the bright optical emission arises from the eastern half of
the HI disk. The velocity field of the galaxy is however quite regular 
and shows a systematic large scale pattern, consistent with the rotational
motion. The  rotation curve for the galaxy shows a peak (inclination corrected)
rotation velocity of only $\sim 8$~km~sec$^{-1}$, comparable to the
random motions in  the HI  gas. After correcting for the 
dynamical support provided by random motions (the ``asymmetric drift'' 
correction), we find the corrected peak rotation velocity of 
$\sim 16.0$~km~sec$^{-1}$. Mass modeling of the corrected rotation curve shows
that the kinematics of DDO210  can be well fit with either a modified isothermal 
halo (with a central density $\rho_0 \sim 29\times10^{-3}$ $M_\odot$ pc$^{-3}$ 
for a stellar mass to light ratio of 3.4) or an NFW halo. In the case of the NFW 
halo however, a good fit is obtained for a wide range of  parameters; the halo 
parameters could not be uniquely determined from the fit. Density profiles 
with inner slope steeper than $\sim 1.2$ however  provide a poor fit to the data.
Finally, the rotation curve derived using MOND also provides a reasonable fit to
the observed rotation curve.
\keywords{Galaxies: evolution --
          galaxies: dwarf --
          galaxies: kinematics and dynamics --
          galaxies: individual: DDO210
          radio lines: galaxies}
}
\maketitle

\section{Introduction}
\label{intro}

Numerical simulations of hierarchical galaxy formation models such as
the CDM model predict a ``universal'' cusped density core for the dark matter halos 
of galaxies (e.g. Navarro et al. 1996). A cusped density core corresponds to a steeply
rising rotation curve. The observed kinematics of galaxies can hence be used to test 
such numerical models of galaxy formation. Dwarf low surface brightness (LSB), galaxies 
are best suited for such a test, since they, unlike larger galaxies, are known to 
be dark matter dominant. In large spiral galaxies, both gas and stars make significant
contributions to the total mass, particularly in the inner regions of the galaxy. 
Since the exact contribution of the baryonic material to the total mass of the 
galaxy depends on the unknown mass to light ratio of the stellar disk, it is 
generally difficult to unambiguously determine the density profile of the dark matter 
halo in the central regions of large spirals. In dwarf LSB galaxies on the other hand, 
since the stellar disk is generally dynamically unimportant, the central halo density 
can be much better constrained. Interestingly, the observed rotation curves of dwarf 
galaxies generally indicate that their dark matter halos have constant density 
cores (e.g. Weldrake et al. 2003) unlike the cusped density cores predicted in 
numerical simulations. Another prediction of the numerical simulations is that the 
density of the dark matter halo is related to the background density at the time of 
the halo formation. Since the smallest galaxies form first in such models, these galaxies
are expected to have the largest halo densities.  The determination of the shapes 
and characteristic densities of the dark matter halos of the faintest dwarf galaxies
is hence a particularly interesting problem. However, a major stumbling block
in such programs is that it is currently controversial whether very faint 
dwarf irregular galaxies show systematic rotation or not. Lo et al. (1993),
in a study of the kinematics a sample of nine faint dwarfs (with M$_B\sim-9.0$ 
to  M$_B\sim-14.0$) found that most of them were characterized by chaotic
velocity fields.  However, as pointed out by ~\cite{skillman96}, Lo et al.'s observations
had limited sensitivity to faint extended emission which is likely to have led to
an underestimation of the rotation velocities. Further, from  a recent high sensitivity and
high  velocity resolution study of the dwarf irregular galaxy, Camelopardalis~B 
(Cam~B),~\cite{begum03} found that inspite of being very faint ($M_B\sim -10.8$), the 
galaxy shows systematic rotation.  Do all faint dwarf irregular galaxies have a 
rotating HI disk or is Cam~B  a special case? What is the dark matter distribution in 
these very faint galaxies? In this paper we discuss these questions in the specific
context of the local group dwarf galaxy DDO210. 

	DDO210, the faintest known ($M_B\sim-10.6$) gas rich dwarf galaxy in our local
group, was discovered by van~den~Berg (1959)  and later detected in  an HI 21 cm survey
by Fisher and Tully (1975). Fisher and Tully (1979) assigned a distance of 0.7~Mpc to it, 
based on its proximity to NGC~6822 both on the sky and in velocity. On the other hand, 
~\cite{greggio93}, based on the colour-magnitude (C-M) diagram of DDO210, estimated its
distance to be  2.5~Mpc. However, recent distance estimates for DDO210 give distances  
closer to the original estimate of Fisher and Tully (1979). \cite{lee99}, based on  the 
I magnitude of the tip of the red giant branch, estimated the distance to DD210 to be 
950$\pm$50~kpc. This estimate is in  excellent agreement with the value of  940$\pm$40~kpc 
derived recently by~\cite{karachentesv02} using HST observations. At this distance,
DDO210 would be a member of the local group. 

DDO210 is classified  as a dIr/dSph or ``transition galaxy'', with properties intermediate
between dwarf irregulars and dwarf spheroidals (Mateo 1998). For example, in spite of 
containing a significant amount of neutral gas, DDO210 shows no signs of ongoing star 
formation. H$\alpha$ imaging detected a single source of line emission in the galaxy; 
however follow up observations of this emission suggests that it does not arise in a 
normal HII region, but probably comes from dense outflowing material from an evolved star
(van Zee et al. 1997). Consistent with the lack of ongoing starformation, the CM diagram
of DDO210 shows that the brightest stars in the galaxy  are the faintest among the
brightest stars in all known dwarf irregular galaxies of our local group (Lee et al. 1999).  

Prior to this work, there have  been two  HI interferometric studies of DDO210, both 
of which used the VLA.  This galaxy was a  part of the sample  of ~\cite{lo93} discussed 
above, and has also been recently re-observed with a  high velocity resolution in the Cs 
array (Young et al. 2003). Although~\cite{young03} noted that DD210 showed a systematic 
large scale  velocity gradient, no attempt was  made to derive a rotation curve for the 
galaxy, since their study focused mainly on the local connection between the ISM and 
star formation . 

	We present here deep, high velocity resolution ($\sim 1.6$~km/s) Giant Meterwave 
Radio Telescope (GMRT) observations of the HI emission from DDO210 and use them to study 
the  kinematics of the neutral gas in this galaxy. The rest of the paper is divided as 
follows. The GMRT observations are detailed in Sect.~\ref{sec:obs}, while the results are 
presented in discussed in Sect.~\ref{sec:res}. Throughout this paper we take the distance
to DDO210 to be 1.0 Mpc, and hence its absolute magnitude to be $M_B \sim -10.6$. 

\section{Observations}
\label{sec:obs}     
    
    The GMRT observations of DDO210 were conducted from 13$-$15 July 
2002. The setup for the  observations is given  in Table~\ref{tab:obs}.
Absolute flux calibration was done using scans on the standard calibrators
3C48 and  3C286,  one of which was observed at the start and end of 
each observing run. Phase calibration was done using the VLA calibrator
source 2008-068 which was observed once every 30 minutes. Bandpass calibration 
was done in the standard way using 3C286. 

     The data were reduced using standard tasks in classic AIPS.  For each run,
bad visibility points were edited out, after which the data were calibrated.
Calibrated data for all runs was combined using DBCON. The GMRT does not
do online doppler tracking -- any required doppler shifts have to be applied
during the offline analysis. However since the differential doppler shift
over our observing interval is much less than the channel width, there was
no need to apply an offline correction.

     The GMRT has a hybrid configuration (Swarup et al. 1991) with 14 of its
30 antennas located in a central compact array with size $\approx$ 1 km 
($\approx$ 5 k$\lambda$ at 21cm) and  the remaining antennas distributed 
in a roughly ``Y'' shaped configuration, giving a maximum baseline length 
of $\approx$ 25 km ($\approx$ 120 k$\lambda$ at 21 cm). The baselines 
obtained from antennas in the central square are similar in length to 
those of the ``D'' array of the VLA while the baselines between the arm 
antennas are comparable in length to the ``B'' array of the VLA. A single 
observation with the GMRT hence yields information on both large and small
angular scales. Data cubes were therefore made at various (u,v) ranges, 
including 0$-$5 k$\lambda$, 0$-$10 k$\lambda$, 0$-15$ k$\lambda$ and
0$-$20 k$\lambda$ using uniform weighting.  At each (u,v) range,  
a circularly symmetric gaussian taper with a FWHM equal to 80\% 
of the (u,v) range was applied inorder to reduce the sidelobes of 
the synthesized beam. The angular resolutions obtained for the various
(u,v) ranges listed above were 44$''\times37''$, 29$''\times23''$, 
20$''\times15''$ and 12$''\times11''$ respectively.
All the data cubes were deconvolved using the the AIPS task IMAGR. 

    The HI emission  from  DDO210 spanned  22 channels of  the spectral cube. 
A continuum image was made using the average of remaining line free
channels. No continuum was detected from the disk of DDO210 to a $3\sigma$
flux limit  of 1.2~mJy/Bm (for a beam size of $30^{''}\times23^{''}$).
We also checked for the presence of any compact continuum sources in 
the disk of DDO210 by making a high resolution ($3.8^{''}\times3.0^{''}$) 
map --  no sources associated with the disk of DDO210 were detected down
to a $3\sigma$ limit of 0.5~mJy/Bm.

The line profiles were examined at various locations in the galaxy and were found 
(to zeroth order) to be symmetric and single peaked. Although double gaussians or 
Gauss-Hermite polynomials provide a better fit to the data (particularly in the higher
column density regions), the  peak velocity of the profile in these regions matches 
(within the errorbars) with the intensity weighted mean velocity.
Since we are interested here mainly in the systematic velocities, moment
maps provide an adequate description of the data.  Moment maps  were
therefore made from the data cubes using  the AIPS task MOMNT. To obtain the moment
maps, lines of sight with a low signal to noise ratio were excluded by
applying a cutoff at the $3\sigma$ level, ($\sigma$ being the rms noise level
in a line free channel), after smoothing in velocity (using boxcar
smoothing three channels wide) and position (using a gaussian with
FWHM $\sim 2$ times that of the synthesized beam). Maps of the velocity 
field and velocity dispersion were also made in GIPSY using  single Gaussian 
fits to the individual profiles. The velocity field produced by gaussian 
fitting is in reasonable agreement with that obtained from moment analysis.
However, the moment2 map systematically underestimates the velocity dispersion.
This can be understood as the effect of the thresholding algorithm used by the MOMNT
task to identify the regions with signal. The velocity dispersion from gaussian 
fitting to the lines profiles  was found to be $\sim$ 6.5 km sec$^{-1}$ and 
showed  no measurable variation across the galaxy. This  lack of substantial 
variation of $\sigma$  across DDO210 is typical of dwarf irregular galaxies, 
although the observed velocity dispersion for DDO210 is lower than the typical 
value observed in  dwarf irregular galaxies (e.g. \cite{lake90,skillman88}) . 
A lower value of HI dispersion for DDO210 was also noted by ~\cite{young03}.

\begin{table}
\caption{Parameters of the GMRT observations}
\label{tab:obs}
\vskip 0.1in
\begin{tabular}{ll}
\hline
Parameters& Value \\
\hline
\hline
RA(2000) & 20$^h$46$^m$53.0$^s$\\
Declination(2000) &  $-{12}^{\circ} 50' 57''$\\
Central velocity (heliocentric) & $-$139.0 km sec$^{-1}$\\
Date of observations & 13$-$15 July 2002\\
Time on source & 16 hrs\\
Total bandwidth & 1.0 MHz\\
Number of channels & 128\\
Channel separation & 1.65 km sec$^{-1}$\\
FWHM of synthesized beam  & 44$''\times37''$, 29$''\times23''$,\\ 
                          & 20$''\times15''$, 12$''\times11''$ \\
RMS noise per channel & 2.2~mJy, 1.8 mJy, 1.6 mJy\\ 
                           &1.4 mJy\\
\hline
\end{tabular}
\end{table}

\section{Results and Discussion}
\label{sec:res}
\subsection{HI distribution}
\label{ssec:HI_dis}

\begin{figure}[h!]
\epsfig{file=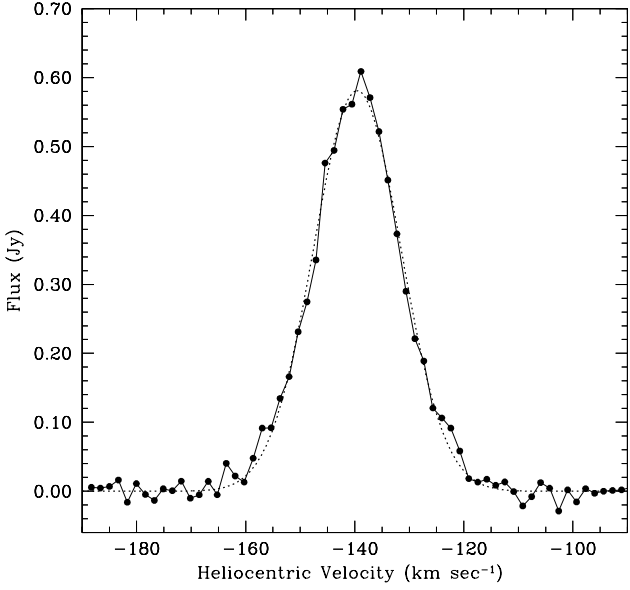,width=3.6in}
\caption{ HI profile for DDO210 obtained from 44$^{''}\times37^{''}$ data cube.
          The channel separation is $1.65$~\kms. Integration of profile 
          gives a flux integral of  $12.1$~Jy \kms and an HI mass of
          $2.8\times{10}^{6} M_\odot$. The dashed line shows 
          a gaussian fit to the profile.
         }
\label{fig:HI_spec}
\end{figure}

   The global HI emission profile of DDO210, obtained from 44$''\times37''$ 
data cube, is shown in Fig.~\ref{fig:HI_spec}. A Gaussian fit to the 
profile gives a central velocity (heliocentric) of  $-139.5 \pm 2.0$~\kms.
The integrated flux is $12.1\pm1.2$~Jy~\kms. These are in good agreement
with the values of $-140 \pm 2.0 $~\kms and  $11.5\pm1.2$~Jy~\kms obtained
from single dish observations (Huchtmeier \& Richter 1986). A good agreement
between the GMRT flux and the single dish flux shows that no flux was
missed because of the missing short spacings in the interferometric
observations. The velocity width at 50 \% level  of peak emission
($\Delta V_{50}$) is found to be  $19.1 \pm 1$~\kms, which again
is in excellent agreement with the $\Delta V_{50}$ value of 19.0~\kms from 
the single dish observations. The HI mass obtained from the integrated profile 
(taking the  distance to the galaxy to be 1.0~Mpc) is $2.8\pm0.3 \times{10}^{6} M_\odot$, 
and the $M_{\rm{HI}}/L_{\rm{B}}$ ratio is found to be $\sim 1.0$ in solar units.

\begin{figure}[h!]
\epsfig{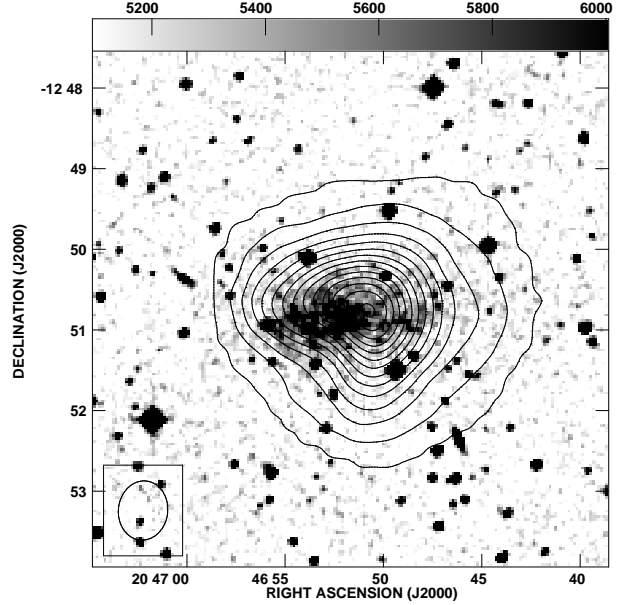}
\caption{The optical DSS image of DDO210 (greyscales) with 
         the GMRT 44$^{''}\times37^{''}$  resolution integrated HI
         emission map (contours) overlayed. The contour levels are 0.01, 0.151, 0.293, 0.434,
         0.576, 0.717, 0.854, 0.999, 1.141, 1.283, 1.424, 1.565, 1.797, 1.848 Jy/Bm~\kms} 
\label{fig:ov}
\end{figure}

\begin{figure}[h!]
\epsfig{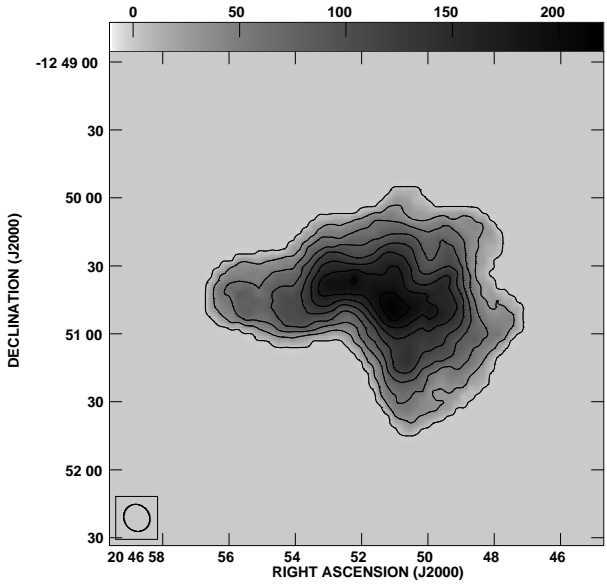}
\caption{ Integrated HI emission map of DDO210 (grey scale and contours) at 12$^{''}\times11^{''}$ resolution.
         The contour levels are 0.001, 0.154, 0.293, 0.434, 0.576, 0.717, 0.858, 0.999, 1.141 Jy/Bm~\kms }
\label{fig:ov_20}
\end{figure}

 Fig.~\ref{fig:ov} shows the integrated HI emission from DDO210 at 
44$''\times37''$ resolution, overlayed on the digitised sky survey (DSS) 
image.  The HI isodensity contours are elongated in eastern and 
southern half of the galaxy i.e. the density is  enhanced in 
these directions.  These density enhancements are  highlighted 
in Fig.~\ref{fig:ov_20} which shows the integrated HI emission at high resolution
(12$''\times11''$). Note that the faint extended emission seen in the low resolution 
map is resolved out in this image. 

  The optical emission in the galaxy shows two components, a central bright 
compact component and an outer faint extended component; both elongated in the 
east-west direction (Lee et al. 1999). However, the centers for the two components do not 
overlap; the bright component lies more eastwards than the center of extended component.
From a deep C-M diagram of DDO210, ~\cite{tolstoy00} found a number of faint stars older
than several Gyr in the galaxy.  ~\cite{vanzee00} estimated an inclination of 68 degrees 
for DDO210  from her observations.  \cite{lee99}, based on deeper observations, estimated 
the inclination of the faint extended component to be $\sim$60 degrees, but noted that
the presence of several bright foreground stars makes this estimate of inclination very
uncertain.

From a visual inspection of the overlay (Fig.~\ref{fig:ov}), the eastern  HI density 
enhancements seem to correlate with the optical emission in the galaxy. As can be 
seen the bright component of the optical emission is not centered on the HI emission,
but is instead offset to the east. On the other hand,  no optical emission is
seen along the southern  HI density enhancement.  As a further check for optical 
emission associated with the southern HI density enhancement, we computed number 
counts of stars in that region using the deep optical image obtained by~\cite{tolstoy00}. 
No significant increase in the star counts from the background was detected.

\begin{figure}[]
\epsfig{file=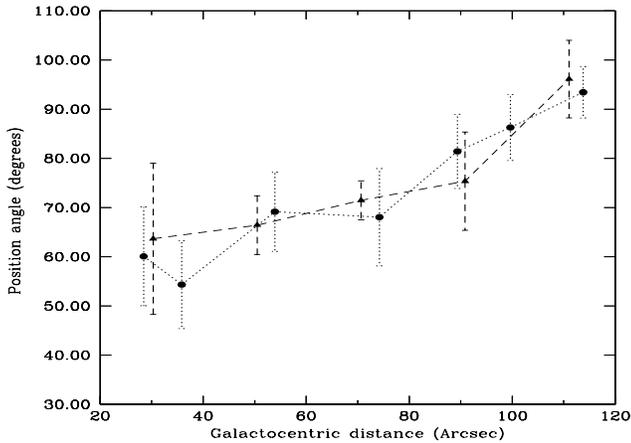,width=3.4in,height=2.6in}
\caption{ The variation of position angle (PA) with the galactocentric
       distance. Circles represent the morphological PA obtained from ellipse fit 
       to the HI distribution at 44$^{''}\times37^{''}$  resolution. 
       Triangles represent the kinematical PA derived from 44$^{''}\times37^{''}$ 
        resolution  velocity field.       
  }
\label{fig:pa}
\end{figure}

As seen in the Fig.~\ref{fig:ov},  the  faint extended emission  in the lower 
resolution HI distribution is less distorted by the presence of the HI density 
enhancements. Hence, an estimate of  the morphological center, position angle (PA)
and inclination of the galaxy  was obtained by fitting elliptical annuli to 
the  44$''\times37''$ resolution HI distribution. The  estimated HI 
morphological center lies close to the center of the faint extended optical 
component (but as noted above is offset from the center of the bright optical
emission). The variation of the fitted morphological  PA  with the 
galactocentric radius is given in Fig.~\ref{fig:pa}. The  inclination 
estimated from the  outer HI contours (assuming the intrinsic shape
of the HI disk to be circular) is 27$\pm$7.0~degrees. The 
inclination  in the inner regions of the galaxy  could not be constrained 
reliably due to the significant distortion in the HI contours. For the same 
reason, ellipse fitting to the higher resolution HI data (where the smooth emission
is resolved out) is not reliable.

\begin{figure}[h!]
\epsfig{file=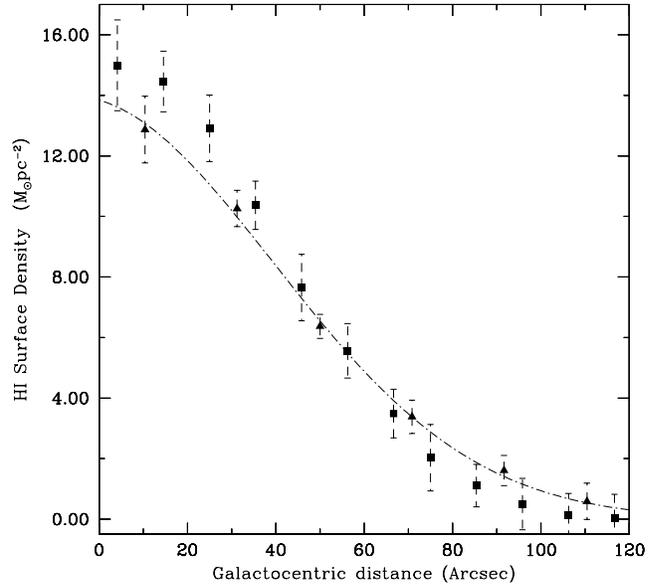,width=3.5in}
\caption{
        The HI surface density profile derived from the HI distribution
        at 29$^{''}\times23^{''}$ (squares) and 44$^{''}\times37^{''}$
         (triangles) resolution. A gaussian fit to 44$^{''}\times37^{''}$
        HI distribution is shown superimposed. 
        }
\label{fig:smd}
\end{figure}

The inclination derived from the HI distribution is significantly different from the 
optical inclination of the galaxy. It is likely that the  optical emission does not
arise from an axi-symmetric disk, but instead  that the stars are concentrated in
a patchy elongated region in DDO210. Hence, in the absence of any other reliable estimate,
the value of inclination obtained from the outer HI contours was assigned to the whole
galaxy. The deprojected HI radial surface density profiles  were then obtained by 
averaging over elliptical annuli in the plane of the galaxy. The profiles derived from
the 44$''\times37''$ and  29$''\times23''$  resolution HI distributions are  given 
in  Fig.~\ref{fig:smd}. As can be seen, the two distributions are in reasonable
agreement. The flux integral estimated from the  29$''\times23''$ resolution HI 
moment~0 map was found to be  $\sim$11\% smaller than that  estimated from the 
44$''\times37''$ resolution HI distribution. For the rest of the analysis, we will 
use the 44$''\times37''$ HI profile. This  profile is  well represented by a  
gaussian, i.e. we have

\begin{equation}
\Sigma_{\rm{HI}}(r)=\Sigma_0\times e^{-(r-c)^2/2r^2_0}
\label{eqn:hisb}
\end{equation}
with  $\Sigma_0$=13.9$\pm$1.0~M$_{\odot}pc^{-2}$, c=$-4.0''$ ($\sim$~0.02kpc)
 and $r_0=46''$ ($\sim$~0.22kpc).

\begin{figure}[]
\epsfig{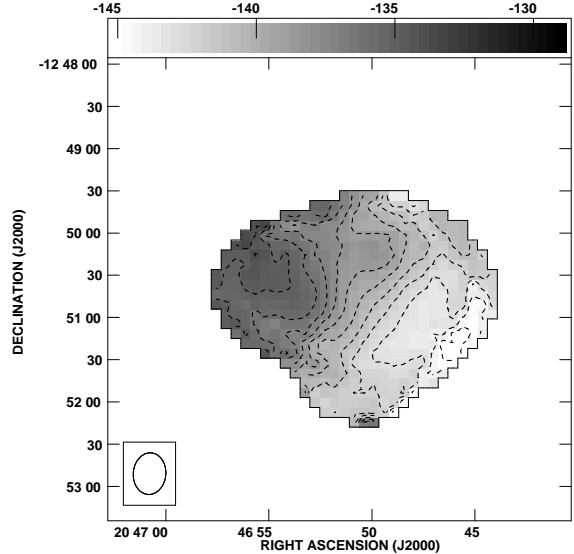}
\caption{The HI velocity field of DDO210 at 29$^{''}\times 23^{''}$ 
          resolution. The contours are in steps of 1~km~sec$^{-1}$ and
          range from $-$145.0~km sec$^{-1}$ to $-$133.0~km~sec$^{-1}$. 
        }
\label{fig:mom1}
\end{figure}

\subsection{HI Kinematics}
\label{ssec:HI_Kin}

	 The velocity field of DDO210  derived from the moment analysis 
of 29$''\times 23''$ resolution data cube is shown in Fig.~\ref{fig:mom1}. 
The velocity field is regular  and a systematic velocity gradient is seen 
across the galaxy. Our velocity field differs significantly from the velocity field
derived by  ~\cite{lo93}. The systematic pattern seen in our velocity field is, to 
zeroth order, consistent with that expected from rotation. On the other hand, the 
velocity field derived by ~\cite{lo93} (based on a coarser  velocity resolution of 
$\sim$ 6 km s$^{-1}$) is chaotic.  This difference in the observed kinematics 
suggests that high velocity resolution and high sensitivity is  crucial in  
determining the systematic gradients in the velocity field of faint galaxies
like DDO210. A similar systematic velocity pattern was also found in the recent
 high velocity resolution study of DDO210 by \cite{young03}, however, those
authors concentrated on trying to determine the physical conditions of the
emitting gas and not the large scale kinematics of the galaxy.

The velocity field in Fig.~\ref{fig:mom1} also shows  signatures  of both
warping and kinematical lopsidedness (\cite{swaters99}), i.e. the kinematical major axis 
is not a straight line but shows twists and the isovelocity contours in the eastern half
of the galaxy are more closed than those in the western half. Both of these kinematical
peculiarities  become more prominent in  the higher spatial resolution velocity field 
(see Fig.~\ref{fig:model}). While the density enhancements in the HI also become more 
prominent at higher spatial resolutions, there does not seem to be any particular 
correlation between the HI density enhancements and  these kinematical peculiarities.

\subsection{HI Rotation Curve}
\label{ssec:rotcur}

\begin{figure}[]
\rotatebox{-90}{\epsfig{file=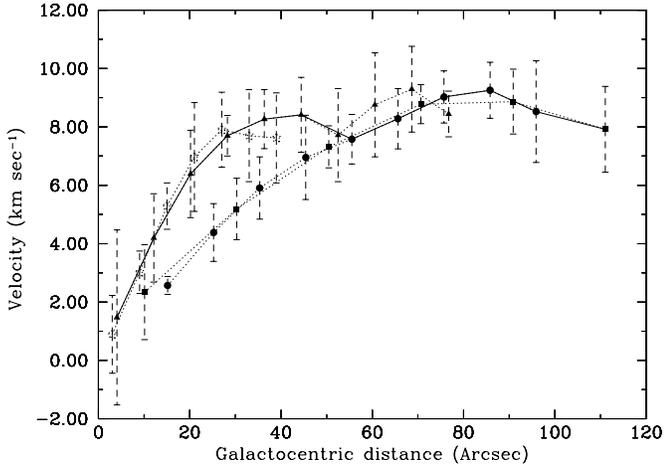,width=2.7in}}
\caption{
The rotation curves derived from the intensity
        weighted velocity field at various resolutions. 
Crosses, triangles, circles and squares
        show the rotation velocity derived from the 12$^{''}\times11^{''}$,20$^{''}\times15^{''}$, 
        29$^{''}\times 23^{''}$ and 44$^{''}\times37^{''}$
        resolution  respectively. 
The adopted hybrid  rotation curve is shown by a solid line. 
  }
\label{fig:vrot}
\end{figure}

Given the lack of correlation between the density enhancements in the HI distribution
and the global kinematical peculiarities in DDO210, it would be reasonable to assume
that these density enhancements follow the same kinematics as that of the more diffuse
extended emission. Under the further assumption that the kinematical peculiarities
noted above are not important (to the zeroth order; see also Sect.~\ref{ssec:discuss}),
rotation curves were derived by fitting the usual tilted ring model to the HI velocity
fields at various resolutions (including 44$''\times37''$, 29$''\times 23''$, 
20$''\times15''$ and 12$''\times11''$) using the GIPSY task ROTCUR.

First, the kinematical center and the systemic velocity (V$_{\rm sys}$)  of DDO210 were 
obtained from a global fit to the velocity fields at various resolutions. The values 
of V$_{\rm sys}$ derived from the velocity fields matched within the errorbars and also 
matched with the value obtained from a gaussian fit to the  global HI profile. The  
kinematical center derived from a  global fit to the velocity fields at 
44$''\times37'' $ and  29$''\times 23''$  resolution matched within the 
errorbars and also matched with the morphological center derived from the 
44$''\times37''$ resolution HI distribution. However, because of  distorted 
morphology of DDO210 at high spatial resolutions, an attempt to derive the kinematical
center from a global fit to the velocity fields at these resolutions did not yield
reliable results. Hence, the center for the higher resolution velocity fields was
fixed to a value obtained from the  lower resolution velocity fields. 

Keeping V$_{\rm sys}$ and the kinematical center fixed to the values obtained from the
global fit, the position angle (PA) of the galaxy was derived, using tilted ring model, 
by breaking up the galaxy into elliptical annuli (each of width half that of the 
synthesized beam). The variation of the derived kinematical PA with the galactocentric 
radius for the 44$''\times37''$ resolution velocity field is given in the 
Fig.~\ref{fig:pa}. The value of  kinematical PA at all galactocentric radii matched 
with the morphological PA estimated  from the 44$''\times37''$ resolution HI 
distribution.  Attempts to derive the kinematical position angle at the higher
spatial resolutions did not yield reliable estimates. Similarly, attempts to derive 
the kinematical inclination of DDO210 did not give reliable results at any spatial
resolution. Hence, in the absence of any other reliable estimate, the kinematical 
inclination of the HI disk was fixed to the value obtained from the  HI morphology. 
Finally, the rotation curve of the galaxy was computed  keeping all  parameters, except 
the circular velocity V$_c$, in each elliptical annuli fixed.

Fig.~\ref{fig:vrot} shows the rotation curves derived  from  the different 
spatial resolution velocity fields. As can be seen, the 20$''\times15''$ 
and 12$''\times11''$  resolution velocity fields lead to the  rotation curves that are
significantly steeper in the inner regions of the galaxy  than those computed from
the lower resolution 44$''\times37''$ and 29$''\times 23''$ velocity fields;
this could be a result of beam smearing. On the other hand,  the rotation curves from 
each  resolution agree in the outer regions of the galaxy. This suggests that the 
effect of beam smearing is  significant only in the inner regions of the galaxy, where 
the  rotation velocity is increasing with the galactocentric radius. Further, the rotation
curves derived from the 12$''\times11''$  resolution velocity field agrees with that 
obtained from the  20$''\times15''$ resolution data suggesting that beam smearing 
effects are no longer significant at such high resolutions (which corresponding to 
a linear scale of $\sim$~70pc). The final rotation curve that we adopt for the
rest of the analysis is shown as a solid line in Fig.~\ref{fig:vrot}. The 
rotation velocities derived from the 20$''\times15''$ resolution velocity 
field are  used in the inner regions of the galaxy (upto $\sim$~50$''$),  
while, the outer points are taken from  the 29$''\times 23''$  and the
44$''\times37''$ resolution data.

\begin{figure*}[th!]
\epsfig{file=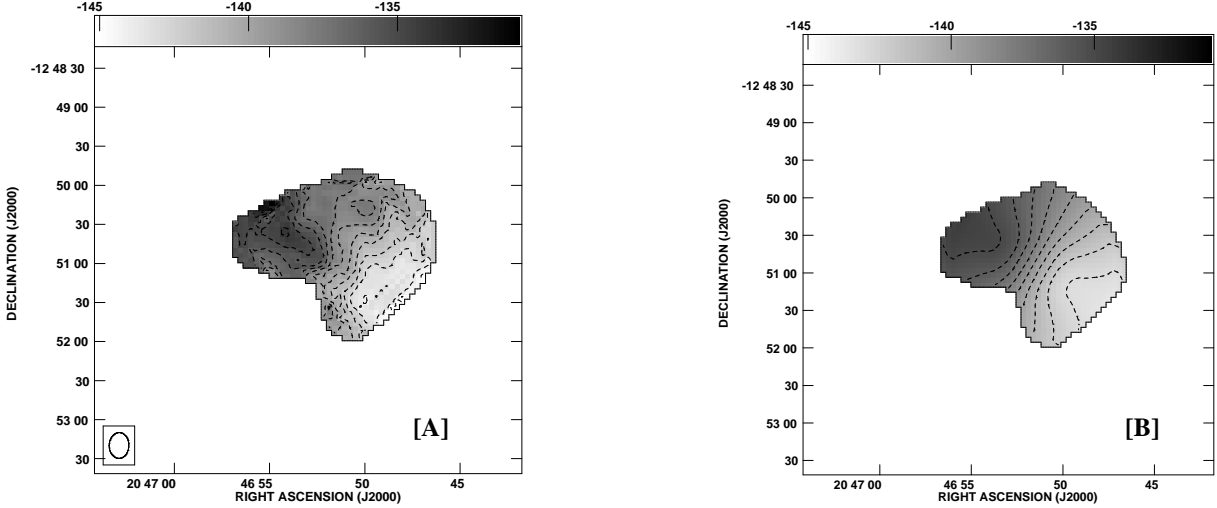,width=3.0in, angle=-90.0}
\caption{\textbf{[A]} The observed  velocity field of DDO210  at  20$''\times15''$ 
arcsec resolution. \textbf{[B]}The model velocity field derived from the rotation 
curve at 20$''\times15''$ resolution. The contours 
    are in steps of 1~km~sec$^{-1}$ and range from -144.0~km sec$^{-1}$ 
    to -134.0~km~sec$^{-1}$.
        }
\label{fig:model}
\end{figure*}

As discussed above, the rotation curves have been derived under the assumption that 
the HI density enhancements follow the same kinematics as that of the extended emission.
The validity  of this assumption could be tested by checking whether the derived 
rotation curves can reproduce the observed kinematics in the galaxy. Hence, model
velocity fields were made at    44$''\times37''$, 29$''\times 23''$ and 
20$''\times15''$ resolutions using the  rotation curves and other kinematical 
parameters, derived from each resolution with a tilted ring model,  using the task 
VELFI in GIPSY. The model velocity field at 20$''\times15''$ resolution  is shown
in ~Fig.~\ref{fig:model}. As can be seen from the figure,
the model provides a  reasonable match to the observed velocity field. Note however
that the model fails to reproduce the kinematical lopsidedness and the twist in the 
kinematical major axis. This is to be expected, since neither of these two features
was incorporated in the model. A similar match between the observed and the model 
velocity fields was also  found for the two lower resolution velocity fields. 

The maximum rotation velocity (inclination corrected) of  DDO210 is $\sim$ 8.0~\kms, 
which is comparable to the velocity dispersion observed in the HI gas, i.e. random 
motion in the gas provides significant dynamical support to the HI disk. Since the 
pressure support contributes significantly to the dynamics of DDO210, the observed 
rotation velocities underestimate the total dynamical mass. The rotation velocity 
hence has to be corrected for the pressure support before one can derive a mass model
for the galaxy; this correction (generally called the ``asymmetric drift'' correction)
is given by:

\begin{equation}
v^2_{\rm{c}}=v^2_{\rm{o}} - r\times{\sigma}^2\bigl[\frac{\rm d}{\rm dr}(\ln{\Sigma_{\rm{HI}}})+\frac{\rm d}{\rm dr}(\ln{\sigma}^2)-\frac{\rm d}{\rm dr}(\ln{2h_z})\bigr],
\label{eqn:ad}
\end{equation}

where $v_{\rm{c}}$ is the corrected circular velocity, $v_{\rm{o}}$ is the observed rotation velocity,
$\sigma$ is the velocity dispersion, and $h_z$ is the scale height of the disk. Strictly 
speaking, asymmetric drift corrections are applicable to collisionless stellar systems 
for which the magnitude of the random motions is much smaller than that of the rotation
velocity. However, it is often used even for gaseous disks, where the assumption
being made is that the pressure support can be approximated as the gas density times the
square of the random velocity. In the absence of any measurement for $h_z$ for DDO210, 
we have assumed $d(\ln(h_z))/dr=0$ (i.e. that the scale height does not change with
radius). Also, using  the fact that $\sigma$ is constant across the galaxy, we get:  

\begin{equation}
v^2_{\rm{c}}=v^2_{\rm{o}} - r\times{\sigma}^2\bigl[\frac{\rm d}{\rm dr}(\ln{\Sigma_{\rm{HI}}})\bigr].
\label{eqn:adrift}
\end{equation}
Using the fitted Gaussian profile to the radial surface density distribution,
(see eqn~\ref{eqn:hisb}) we obtain
\begin{equation}
v^2_{\rm{c}}=v^2_{\rm{o}} + r(r-c)\times\sigma^2/r^2_0.
\label{eqn:corr_curve}
\end{equation}

 The observed HI velocity dispersion was corrected for the instrumental broadening
as well as  for the  broadening due to the  velocity gradient over the finite 
size of the beam. The applied correction is 

\begin{figure}[h!]
\epsfig{file=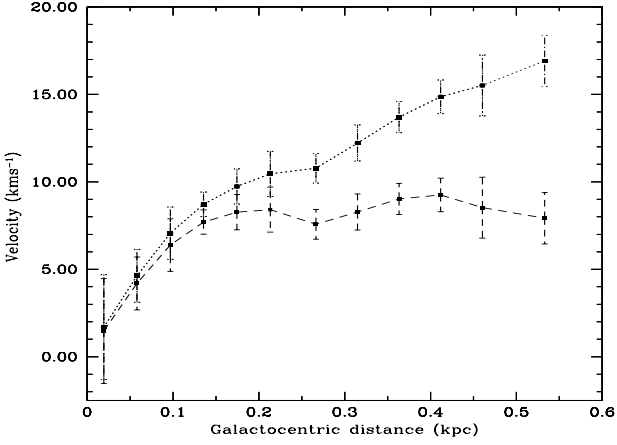,width=3.4in,height=2.6in}
\caption{  The hybrid rotation curve (dashes)  and the rotation curve after applying
          the asymmetric drift correction (dots). 
  }
\label{fig:v_asy}
\end{figure}

$\sigma_{\rm true}^{2}=\sigma_{\rm obs}^2-\Delta v^2-
        \frac{1}{2}{b}^2{({\nabla}v_{\rm{o}})}^2$,

where  $\sigma_{\rm true}$ is the true velocity dispersion, $\Delta v$ is the channel width,
$b$ characterizes the beam width (i.e. the beam is assumed to be of form $e^{-x^2/b^2}$)
and $v_{\rm{o}}$ is the observed rotation velocity. After putting the appropriate values in the
above  equation, we get  $\sigma^2_{\rm true} \approx 36.0$~km$^2$~sec$^{-2}$.
Finally, substituting this value back into the Eqn.~(\ref{eqn:corr_curve})
the ``asymmetric drift" corrected curve obtained is given as a dotted line in
Fig.~\ref{fig:v_asy}.

As discussed above, the ``asymmetric drift'' correction was
derived under the assumption  that the disk scale height of DDO210 does not 
change with radius. To quantify the effect of this assumption, the 
correction  was recalculated  assuming a linear increase 
of 100\% in the scale height, from q$_0$=0.25 at  the center to q$_0$=0.5 at 
the edge of the galaxy. The change in the asymmetric drift correction was 
found to be $<1$\%. The assumption that the scale height is constant hence 
have a negligible effect on the derived halo parameters.   

\subsection{Mass Model}
\label{ssec:massmodel}

In this section we use the ``asymmetric drift" corrected rotation curve 
derived in the last section to derive mass models for DDO210.  

As discussed in Sect.~\ref{ssec:HI_dis} the optical emission is not axi-symmetric but
is instead patchy and elongated. For the purpose of mass modeling however, we approximate
the optical emission to be axi-symmetric, but we return to this issue in 
Sect~\ref{ssec:discuss}. Assuming that the optical  emission was axi-symmetric, \cite{lee99}
estimate the B band scale length ($\alpha_B$) of DDO210 to be 35.0${''}$ ($\sim 0.17$~kpc).
This estimate of $\alpha_B$ agrees to within 10\% with the value of 39$^{''}$ estimated 
by~\cite{vanzee00}.  \cite{lee99} also found the colours of the galaxy to 
be nearly constant with the galactocentric radius. Hence, we computed the contribution 
of the stellar mass  to the observed rotation curve by assuming it to be an exponential
disk of constant M/L$_B$ ratio ($\Upsilon_B$) and  with an intrinsic thickness ratio 
(q$_0$) of 0.25. For the vertical density distribution of the stellar disk, we assumed a 
 sech$^2(z/z_0)$  profile, with $z_0$ independent of galactocentric radius. These are 
reasonable assumptions for disk galaxies (e.g. van~der~Kruit \& Searle 1981, 
de~Grijs \& Peletier 1997). In the absence of any prior knowledge of $\Upsilon_B$, 
it was taken as a free parameter in the mass modelling.

   The contribution of the HI mass to the observed rotation velocities was 
calculated using the  HI surface density profile derived from 44$''\times37''$ 
HI distribution. To correct for the mass fraction of Helium, the HI 
mass was scaled by a factor of 1.25. A search for molecular gas in DDO210
gave negative result (Taylor et al. 1998), hence, the  contribution of molecular gas to 
the rotation curve was ignored. We also neglected the contribution of ionized gas, 
if any. Not much is known about the vertical distribution of gas in dwarf irregular 
galaxies,  however there is some evidence of similar vertical density distribution 
of HI and stellar disk (Bottema et al. 1986). Hence, for the HI disk  we again assumed 
q$_0$ of 0.25, with a vertical density distribution profile of sech$^2(z/z_0)$. The 
rotation velocities for the HI and the stellar components were then computed using the 
formulae given by ~\cite{casertano83}. 

                For the dark matter halo we considered two types of density
profiles, viz. the modified isothermal profile and the NFW profile. The modified isothermal
density profile is given by:

$\rho_{\rm iso}(r)=\rho_0[1+{(r/r_{\rm{c}})}^2]^{-1}$,

where, $\rho_0$ is the central density of the halo and  $r_c$ is the core radius.
The corresponding circular velocity is given by:

$v(r)=\sqrt{4\pi G \rho_0 r^2_c\bigl[1-\frac{r}{r_{\rm{c}}}\tan^{-1}(\frac{r}{r_{\rm{c}}}\bigr)]}$.

The NFW halo density profile is given by:

$\rho_{\rm NFW}(r)=\rho_i/[(r/r_s)(1+r/r_s)^2]$,

 where, $r_s$ is the characteristic radius of the halo and $\rho_i$ is the characteristic
density. The circular velocity can be written as:

$v(x)=v_{200} \sqrt{\frac{\ln(1+cx)-cx/(1+cx)}{x[\ln(1+c)-c/(1+c)]}}$,

where,  $c = r_{200}/r_s$, $x = r/r_{200}$; $r_{200}$ is the distance at which the mean
density of the halo is equal to 200 times the critical density and $v_{200}$ is the 
circular velocity at this  radius.

\begin{figure}[h!]
\rotatebox{-90}{\epsfig{file=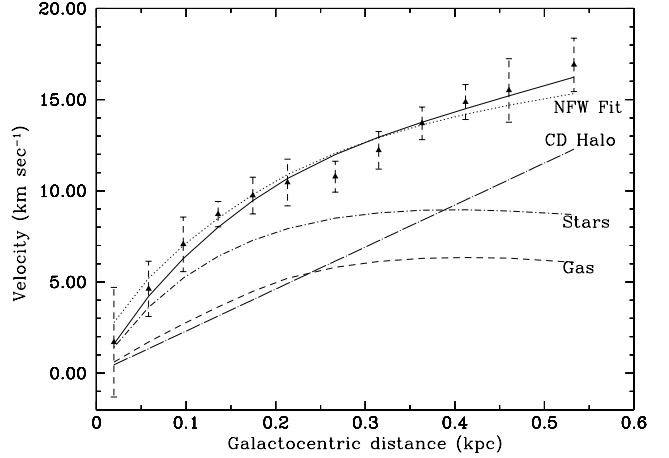,width=2.6in}}
\caption{ Mass models for DDO210 using the corrected rotation curve.
         The points are the observed data. The total mass of gaseous disk (dashed line)
          is $3.6\times10^6 M_\odot$.The stellar disk (short dash dot line) has
          $\Upsilon_B=3.4$, giving a stellar mass of $9.2 \times10^6 M_\odot$. The
          best fit total rotation curve for the constant density halo model is shown as
          a solid line, while the contribution of the halo itself is shown as a
          long dash dot line (the halo density is density 
$\rho_0=29~\times10^{-3} M_\odot$ pc$^{-3}$). 
The best fit total rotation curve for an NFW type halo,
          using $\Upsilon_B=0.5$, c=5.0 and $v_{200}$=38.0~\kms
    is shown as a dotted line. See the text for more details. 
  }
\label{fig:massmodel}
\end{figure}

First, we consider a modified  isothermal dark halo mass model. Since the rotation curve 
keeps on rising till the last measured point (see Fig.~\ref{fig:v_asy}) , the core radius 
$r_c$ could not be constrained. This essentially reduces a modified  isothermal dark halo 
fit into a constant density halo fit. The remaining free parameters are the central 
density of the halo, $\rho_0$, and  the mass to light  ratio of the stellar disk, 
$\Upsilon_B$. 

 Fig.~\ref{fig:massmodel} shows the best fit mass model for a constant density halo. The 
best fit gave $\Upsilon_B$ of 3.4$\pm$0.5 and 
$\rho_0=29\pm5~\times10^{-3}~M_\odot$~pc$^{-3}$. The observed B-V for DDO210 is 
$\sim$0.25 (Lee et al. 1999), which, (from the low metallicity Bruzual $\&$ Charlot SPS 
model using a modified Salpeter IMF, Bell $\&$ de Jong 2001), corresponds to a 
$\Upsilon_B$ of $\sim$ 0.5 . If we adopt this value of  $\Upsilon_B$ then the best fit 
model has $\rho_0=59\pm7~\times10^{-3} M_\odot$ pc$^{-3}$. However, with  this  
$\Upsilon_B$, the fit to the observed rotation curve is much poorer and the model curve  
substantially underestimates the observed rotation velocities at small radii.
The best fit model ($\Upsilon_B=3.4$) gives  the mass of stellar disk to be 
$M_*$=$9.2\times 10^6 M_\odot$.The mass of the gas disk in DDO210  is 
$M_{\rm gas}$=$3.5\times10^6 M_\odot$. From the last measured point of the  observed rotation
curve, we get a total dynamical mass of $M_T$=3.4$\times 10^7 M_\odot$, i.e. at the 
last measured point more than 63\% of the mass  of DDO210 is dark. 

     A similar procedure was also tried using a dark matter halo of the NFW type. 
Keeping $\Upsilon_B$ as a free parameter in the fit gave unphysical results, hence, 
it was kept fixed  to  a more likely value (viz. $\Upsilon_B = 0.5$). We found that an
NFW halo provides a good fit to the data, for a wide range of  values  of $v_{200}$ 
and $c$. The range of parameters which provide acceptable fits are ($v_{200}~\sim $20~\kms,
$c \sim 10$) to ($v_{200}~\sim $500~\kms, $c \sim 0.001$). As an illustration we show in 
Fig.~\ref{fig:massmodel} the best fit rotation  curve for an NFW halo, using 
$\Upsilon_B=0.5, c=5, v_{200}=38$~\kms. As can be seen clearly, the NFW halo also provides 
a good fit to the data. However, the best fit values of the concentration parameter c, 
at any given $v_{200}$ was found to be consistently smaller than the value predicted by
numerical simulations for the  $\Lambda$CDM universe (Marchesini et al. 2002).

  As seen above, both isothermal and NFW halo provide  a good fit to
the observed kinematics of DDO210. Profiles steeper than NFW have
also been proposed by some N-body simulations (e.g. Moore et al. 1999).
In order to check whether such steep profiles  are also consistent
with the data, mass models were fit using a broader family  
of density profiles, viz.

\begin{equation}
\rho(r)=\frac{\rho_0}{{(r/r_0)}^\alpha{[1+{(r/r_0)}^\gamma]}^{(\beta-\alpha)/\gamma}}
\label{eqn:dens}
\end{equation}

\begin{figure}[h!]
\rotatebox{-90}{\epsfig{file=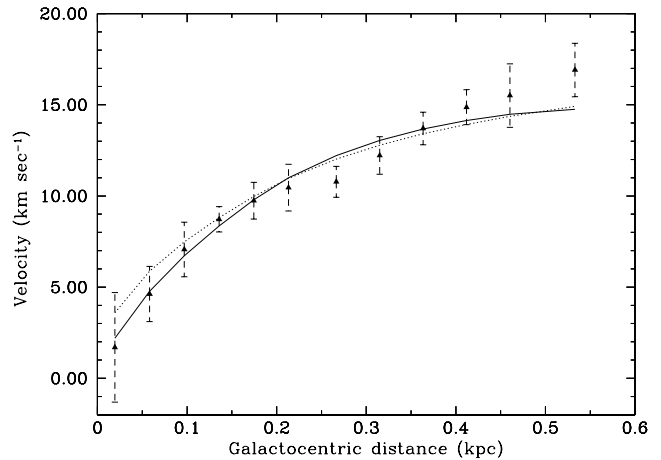,width=2.6in}}
\caption{  The best fit MOND rotation curve (solid line) to the 
         ``asymmetric drift'' corrected rotation velocities (points). 
           Also shown in the figure is the best fit total rotation curve for a  dark 
          halo with $\alpha=1.2$ inner slope (dotted line). 
	  See the text for more details.
  }
\label{fig:mond}
\end{figure}

 The circular velocity corresponding to the above density profile 
(see Kravtsov et al. 1998) is:

\begin{equation}
\rm{V}(r)=V_{\rm{t}}\frac{{(r/r_{\rm{t}})}^g}{{[1+{(r/r_{\rm{t}})}^a]}^{(g+b)/a}}
\end{equation}

where  r$_{\rm{t}}$ and V$_{\rm{t}}$ are the effective ``turnover" radius and 
velocity. The parameter ``g" is related to the inner slope of the 
density profile, $\alpha$  by g$= 1+ \alpha/2.0$, ``b" is the outer 
logarithmic slope of the rotation curve while ``a" determines 
the sharpness of turnover. Fits to the rotation curve with 
all three parameters left free did not converge. Hence, since 
we are primarily concerned with the slope in the inner regions, 
we fixed the parameters b and  a to the values of 
0.34 and 1.5  respectively, (which are the typical values found 
for the rotation curves of dwarf galaxies; Kravtsov et al. 1998),
while r$_{\rm{t}}$ and V$_{\rm{t}}$ were left as free parameters.  
$\Upsilon_B$ was fixed to a value of 0.5, as  suggested by the 
observed colours in the galaxy; this also allows a meaningful 
comparison of  derived mass models for a family of density profiles.  
We found that the reduced $\chi^2$ for the fit continuously increases 
as the profile gets steeper. For comparison, fixing g to 0.5 (corresponding 
to $\alpha=$1.0; NFW profile) gave reduced $\chi^2=$0.4 while g 
of 0.4 (corresponding to $\alpha=$1.2) gave reduced $\chi^2=$0.7.
At the extreme, fixing g to 0.2 ($\alpha=$1.6) (which substantially
over-predicts  the observed rotation velocity at small radii, 
while  underestimates the velocity at large radii) gave reduced 
$\chi^2=$2.5. Note that since there are heuristics involved in 
computing the error bars on $v_c$, it is not possible to 
rigorously translate the minimum $\chi^2$ value into a 
confidence interval for the parameters of the fit. However, 
a lower $\chi^2$ value does imply a better fit (see also 
the discussion in van den Bosh \& Swaters 2001). 
Fig.~\ref{fig:mond} shows the best fit mass model 
with $\alpha=1.2$.

 So far we have assumed that the discrepancy between the dynamical mass,
estimated from the ``asymmetric drift'' corrected rotation curve, and 
the  luminous  mass in DDO210 can be explained by considering 
an extended dark halo around the galaxy. An alternate 
explanation for this  discrepancy is that the dynamics becomes non-Newtonian 
in the limit of low acceleration, i.e. the MOND theory (Milgrom 1983). 
We have also tried fitting the rotation curve using the MOND prescription.
In this fit,  $\Upsilon_B$ and a$_0$ (the critical acceleration
parameter) were taken as  free parameters. Fig.~\ref{fig:mond} 
shows the best fit MOND rotation curve. As can be seen, the 
MOND rotation curve agrees well with the observed curve in the 
inner regions of the galaxy, while  it underestimate the 
observed curve (by up to 2.0 kms$^{-1}$) in the outer regions. 
The best fit model gave $\Upsilon_B$ of 0.4$\pm$0.2 and a$_0$ 
of 1.7$\pm$0.3 (in units of 10$^{-8}$ cms$^{-2}$) with reduced 
$\chi^2=$0.7. The best fit value of $\Upsilon_B$ agrees with  the value
expected from the observed colours in DDO210. Also, the best fit 
value of a$_0$ is consistent with the mean value of a$_0$  
found for other, brighter galaxies (Kent (1987); see also Milgrom 1988). 
DDO210 is the faintest known dwarf irregular galaxy for which the 
MOND prescription provides a reasonably good fit to the observed kinematics.

   Recall that all the above mass models have been derived by  assigning 
an  inclination of 30 degrees (obtained from the outer HI contours, as discuss 
in~Sect.\ref{ssec:HI_dis}). Inorder to  estimate the  effect of using an 
erroneous inclination on the derived halo parameters, mass modeling was 
also tried  with two other values of inclination viz. 60 degrees  (i.e. that 
estimated from the optical isophotes) and 45 degrees (a value 
between the optical and HI inclination).

   The best fit mass model for a constant density halo, using an
inclination of 60 degrees,  gave  $\Upsilon_B$ of 1.4$\pm$0.2 and 
$\rho_0=27\pm2.0~\times10^{-3} M_\odot$ pc$^{-3}$ with reduced $\chi^2=0.3$.
On the other hand, an inclination of 45 degrees gave the  best fit model 
with $\Upsilon_B$ of 1.8$\pm$0.2  and $\rho_0=31\pm2.0~\times10^{-3} M_\odot$ pc$^{-3}$ 
with reduced $\chi^2=0.2$. Recall that an inclination of 30 degrees 
gave the  best fit with $\Upsilon_B$ of 3.4$\pm$0.5 and 
$\rho_0=29\pm5.0~\times10^{-3} M_\odot$ pc$^{-3}$ with 
reduced $\chi^2=0.4$. As can be seen,  the central halo density is 
relatively insensitive to the assumed  inclination. However, the best 
fit $\Upsilon_B$ changes significantly with the assumed inclination. 
 
For the  NFW halo fit, keeping $\Upsilon_B$ as a free parameter in the 
fit gave unphysical results, hence, it was kept  fixed to a value of 
0.5, as suggested by the observed colours in the galaxy. The  best fit NFW 
model with an  inclination of 60 degrees  gave a reduced $\chi^2=4.0$, 
while for 45 degrees a reduced $\chi^2=1.6$ was obtained from the best fit.	
As can be seen, the  NFW halo provides a poor fit to the data 
for higher values of inclination (for comparison, an  inclination  
of 30 degrees gave the  best fit NFW model with a  reduced 
$\chi^2=0.5$). Hence, an  NFW halo can be ruled out for DDO210,  
if the inclination of the galaxy  is higher than 30 degrees.

\subsection{Discussion}
\label{ssec:discuss}

\begin{figure}[h!]
\rotatebox{-90}{\epsfig{file=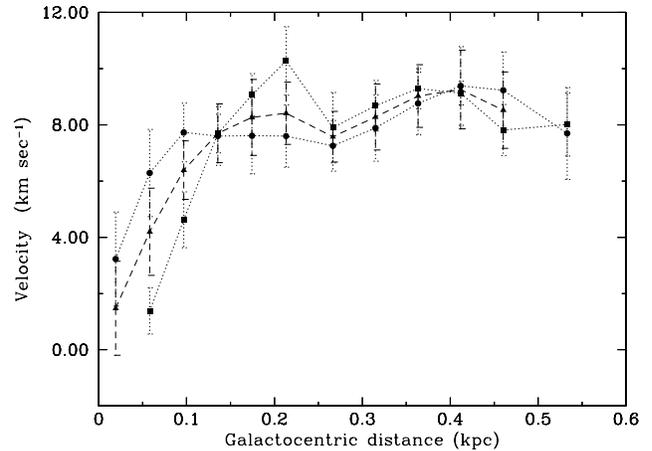,width=2.6in}}
\caption{ Filled triangles show the rotation velocities derived from the whole
         galaxy while circles and squares represent the rotation velocities derived
        from the receding and the approaching side respectively. }
\label{fig:vrotall}
\end{figure}

As discussed in the Sect.~\ref{ssec:HI_Kin}, DDO210 shows signatures of twisting of
the kinematical major axis as well as of kinematical lopsidedness. This may be related
to the lopsidedness seen in the optical disk; the bulk of the optical emission (i.e.
the ``bright component'' discussed in Sect.~\ref{ssec:HI_dis}) lies in the eastern
half of the HI disk. A similar pattern of  kinematical lopsidedness and 
lopsidedness in the optical disk is also seen in other galaxies (e.g. M101, 
Bosma et al. 1981). To quantify the kinematical lopsidedness in the galaxy, rotation 
curves were derived at various spatial resolutions separately for the approaching and 
receding halves of the galaxy. Fig.~\ref{fig:vrotall} shows the hybrid rotation curves 
for the approaching and receding sides as well as for the galaxy as a whole. The rotation
curves obtained from the two  lowest  resolution  velocity fields 
(44$^{''}\times37^{''}$ and 29$^{''}\times 23^{''}$) matched within the errorbars 
with the one derived from the whole galaxy. On the other hand, for the higher resolution
data, the rotation curves were found to be different (at $\sim$~2.0~\kms level)
from the rotation curve derived  from the whole galaxy. While part of this difference
may be due to the kinematical lopsidedness, it is also possible that part of it
arises as a result of the  difference in the sampling of the velocity field between the 
two sides of the galaxy. Because of  the distorted morphology of the HI distribution, the 
distribution of data points available to estimate rotation velocity  are different for
the approaching and receding halves. This effect gets more important at the higher 
spatial resolutions and could also contribute to the observed differences in the
rotation curves. Further, as can be seen in Fig.~\ref{fig:vrotall} the 
difference between the rotation curve derived using the whole galaxy and the rotation curves 
derived separately for the two halves  are small compared to both the error bars on
the average curve as well as the magnitude of the asymmetric drift correction. Ignoring 
the kinematical lopsidedness for the mass modeling is hence probably reasonable.

The derived rotation curve can also be used to  put limits on the dynamical ages  of the 
density enhancements in the HI distribution. The derived rotation curve for the galaxy 
(see Fig.~\ref{fig:vrot}) is flat from a galactocentric radii of 0.2 kpc to 0.5 kpc. 
The time scales required for one rotation in the inner regions of the galaxy is 
$\sim$ 160 Myr, while the rotation period at the edge of the galaxy is $\sim$ 400 Myr. 
Hence, this differential rotation  will wind up these density enhancements on a time 
scale of a  few 100~Myrs. These density enhancements are hence likely to be recently
developed perturbations in the HI disk. The origin of these perturbations is unknown,
and it also surprising that despite their presence and relative dynamical youth there
is not much evidence for star formation in the galaxy.

To conclude, we have presented  deep, high velocity resolution ($\sim 1.6$ km sec$^{-1}$) 
GMRT HI 21cm synthesis images for the faint ($M_B \sim -10.6$) dwarf galaxy 
DDO210. We find that  the HI distribution in the galaxy is not axi-symmetric, but shows 
density enhancements in the eastern and southern parts of the galaxy. The velocity field 
of the galaxy is however regular, and shows a systematic large scale pattern, consistent
with rotational motion. The high velocity resolution ($\sim 1.6$ km sec$^{-1}$) and 
sensitivity of our observations were crucial for determining the true velocity field of
the galaxy. The inclination angle for DDO210 is poorly known. From the observed 
velocity field, we derive a rotation curve for DDO210 by assigning an inclination derived 
from outer HI contours to the whole galaxy. The derived peak rotational velocity is  
comparable to the random motions in  the HI  gas.  After correcting for the dynamical 
support provided by the random motion, we find that the rotation curve of DDO210  can 
be well fit with  either a modified isothermal halo (with a central density $\rho_0 \sim 
29\times10^{-3}$ $M_\odot$ pc$^{-3}$) or an NFW halo. We find that for a constant 
density halo, the central halo density is relatively insensitive to the assumed 
value of inclination. However, the best fit $\Upsilon_B$ changes 
significantly with the assumed inclination. On the other hand, an NFW halo can be 
ruled out for DDO210, if the inclination of the galaxy is higher than $\sim 30$~degrees.
Mass models are also derived  for a family of density profiles steeper than NFW; we find that 
density profiles with inner slopes steeper than  $\sim 1.2$ provide a poor fit to the data.

    As an alternative  to the dark matter hypothesis, we have derived MOND
fits to the rotation curve. We find that the rotation curve predicted by MOND 
provides an acceptable fit to the observed rotation curve. Further, the value of the
critical acceleration a$_0$ of 1.7$\pm$0.3 $\times 10^{-8}$ cms$^{-2}$ is consistent 
with the earlier determinations of the value of this parameter.

\begin{acknowledgements}

        The observations presented in this paper would not have been possible without 
the many years of dedicated effort put in by the GMRT staff in order to build the 
telescope. The GMRT is operated by the National Centerer for Radio Astrophysics of 
the Tata Institute of Fundamental Research. We are grateful to Dr E. Tolstoy for
providing a reduced  VLT image  of DDO210.

\end{acknowledgements}

\end{document}